\documentclass[letterpaper]{JHEP3}

\usepackage{amsmath,amssymb}
\usepackage{graphicx}

\newcommand{\be}{\begin{equation}}
\newcommand{\ee}{\end{equation}}
\newcommand{\ba}{\begin{aligned}}
\newcommand{\ea}{\end{aligned}}
\newcommand{\ben}{\begin{displaymath}}
\newcommand{\een}{\end{displaymath}}
\newcommand{\bea}{\begin{eqnarray}}
\newcommand{\eea}{\end{eqnarray}}
\newcommand{\bean}{\begin{eqnarray*}}
\newcommand{\eean}{\end{eqnarray*}}

\def\e {\varepsilon}

\preprint{Brown-HET-1583}

\DeclareMathOperator{\Str}{Str}

\title{Factorized Tree-level Scattering in $AdS_4 \times \mathbb{CP}^3$}

\author{Chrysostomos~Kalousios, C.~Vergu, Anastasia~Volovich\\
Department of Physics, Brown University, Box 1843, Providence, RI 02912 \\
E-mail: \email{ckalousi@het.brown.edu, Cristian\_Vergu, Anastasia\_Volovich@brown.edu}
}

\abstract{$AdS_4/CFT_3$ duality relating IIA string theory on $AdS_4 \times \mathbb{CP}^3$ to $\mathcal{N}=6$ superconformal Chern-Simons
theory provides an arena for studying aspects of integrability in a new potentially exactly solvable system.  In this paper we explore the tree-level worldsheet scattering for strings on $AdS_4 \times \mathbb{CP}^3.$
We compute all bosonic four-, five- and six-point amplitudes in the gauge-fixed action and demonstrate the absence of particle production.}


\begin{document}

\section{Introduction}

Integrability of both string and gauge theory sides of the $AdS_5/CFT_4$ correspondence plays a very important role in exploring various aspects of the duality.  In particular it leads to an exact solution for certain all-loop quantities in $\mathcal{N}=4$ Yang-Mills theory \cite{Beisert:2006ez, Beisert:2006ib}.

A new example of $AdS_4/CFT_3$ duality relating IIA string theory on $AdS_4 \times \mathbb{CP}^3$ to $\mathcal{N}=6$ superconformal Chern-Simons theory has been proposed by Aharony, Bergman, Jafferis and Maldacena~\cite{Aharony:2008ug} building on earlier work by Bagger, Lambert and Gustavsson~\cite{Bagger:2006sk, Gustavsson:2007vu,   Bagger:2007jr, Bagger:2007vi, Gustavsson:2008dy}.  It provides a new arena for studying aspects of integrability on both sides of the correspondence.  On the string theory side, the worldsheet action for IIA on $AdS_4 \times \mathbb{CP}^3$ has been constructed and its possible integrability has been explored in~\cite{Arutyunov:2008if,   Stefanski:2008ik, Grignani:2008is, Fre:2008qc, Astolfi:2008ji, Bonelli:2008us, Gomis:2008jt,   Grassi:2009yj, Zarembo:2009au}.  On the gauge theory side, two-loop integrability has been explored in~\cite{Gaiotto:2007qi,   Minahan:2008hf, Gaiotto:2008cg, Bak:2008cp, Bak:2008vd,   Zwiebel:2009vb, Minahan:2009te,Papathanasiou:2009en, Ahn:2009zg} where the corresponding spin chain has been constructed.  An exact $S$-matrix has been proposed in~\cite{Gromov:2008qe, Ahn:2008aa,Ahn:2008tv,Gromov:2009tv} and various tests of this new $AdS_4/CFT_3$ duality have been carried out in~\cite{McLoughlin:2008ms, Alday:2008ut, Krishnan:2008zs, Gromov:2008fy, McLoughlin:2008he, Sundin:2008vt, Zarembo:2009au}.

Even though string theory in the $AdS_5 \times {S}^5$ background is classically integrable, explicitly solving worldsheet string theory at finite values of the coupling is a very difficult problem.  Perturbative study of the $S$-matrix of the full string worldsheet in $AdS_5 \times {S}^5$ has been initiated in~\cite{Klose:2006zd}, where worldsheet scattering amplitudes in the light-cone gauge have been calculated to leading order in perturbation theory and the supersymmetry realization on scattering states has been analyzed.  The truncation of the $AdS_5 \times {S}^5$ string action to the near flat space limit and the calculation of its scattering amplitudes  up to two loops have been performed in~\cite{Klose:2007wq, Klose:2007rz, Puletti:2007hq}.

Recently, Zarembo has studied the worldsheet $S$-matrix for the $AdS_4 \times \mathbb{CP}^3$ sigma model
\cite{Zarembo:2009au}.  He proposed a solution to the mismatch between the number of degrees of freedom present in the Bethe ansatz and in the sigma model by showing that, in the sigma model Green functions,
the position of the poles of the heavy modes is at the threshold of producing two light particles and quantum corrections make this pole disappear.  Thus, the massive excitations ``dissolve'' in the continuum of light states, leaving only the light states which fall into a representation of $SU(2|2)$, which is also the symmetry group of the $S$-matrix proposed in~\cite{Ahn:2008aa}. Zarembo also checked that the conjectured exact $S$-matrix of $AdS_4/CFT_3$ duality agrees with the tree-level worldsheet calculations for the bosonic four-point amplitudes.

Arutyunov and Frolov constructed a Lax connection for the $AdS_4 \times \mathbb{CP}^3$ coset sigma model, from which classical integrability follows~\cite{Arutyunov:2008if}.  Their construction of the Lax connection is inspired by the earlier study of the $AdS_5 \times {S}^5$ background in~\cite{Bena:2003wd}.  Integrability implies the presence of higher conservation laws which, in turn, forbid particle production in the scattering process and require $S$-matrix factorization (see~\cite{Zamolodchikov:1978xm}).  In this paper we will check by explicit calculations the absence of particle production in the $AdS_4 \times \mathbb{CP}^3$ sigma model.  We will do this by computing bosonic tree-level scattering amplitudes in the gauge-fixed string action through six points.  The gauge-fixed action contains a free parameter $a$, which we will keep arbitrary throughout the computation.  The $S$-matrix will be a polynomial in $a$ and the absence of particle production implies that all the coefficients of this polynomial in $a$ must vanish.  This imposes stringent constraints on the $a$-dependent parts of the action.

One can construct a formal argument for the classical integrability of the gauge-fixed action.  The initial action is classically integrable and gauge fixing proceeds by imposing a set of first-class constraints which are compatible with the equations of motion and finally eliminating the gauge degrees of freedom by using these constraints. The higher conserved charges should descend to higher conserved charges on the reduced configuration space.  However, the gauge-fixing step is subtle so it is nice to be able to check explicitly that the integrability survives.  This paper provides such an explicit check.

Quantum integrability of strings in $AdS_5 \times S^5$ has been explored in \cite{Berkovits:2004xu}, but
quantum integrability of the $AdS_4 \times \mathbb{CP}^3$ model is less certain because, unlike the $SO(N)$ models, the $\mathbb{CP}^N$ models are known not to be integrable (see~\cite{Abdalla:1980jt,   Abdalla:1982yd}).  Because of this, checking the quantum integrability of this model is an important open problem.  The on-shell tree-level amplitudes computed in this paper could also be important as ingredients in the unitarity method to construct amplitudes at loop level. The problem of quantum integrability perhaps could be addressed in a simpler case, like the reduced near flat space action constructed by Maldacena and Swanson~\cite{Maldacena:2006rv} for the $AdS_5 \times {S}^5$ theory.  However, such a reduced action has not been constructed yet for the $AdS_4 \times \mathbb{CP}^3$ background (see~\cite{Kreuzer:2008vd} for the bosonic part of the action). Although the fermion interactions should be fixed by supersymmetry, it would be interesting to also compute scattering amplitudes with fermions explicitly.  They will of course be very important at loop level.

The paper is organized as follows. In section~\ref{sec:sigma_model} we review the sigma model description for strings in $AdS_4 \times \mathbb{CP}^3$, restricting to the bosonic fields.  In section~\ref{sec:gauge_fixing} we review the gauge fixing procedure, working out the Lagrangian up to sixth order.  In section~\ref{sec:scattering_amplitudes} we present our results for four-, five- and six-point bosonic scattering amplitudes, and demonstrate the absence of particle production.

\section{Sigma model description}
\label{sec:sigma_model}

In this section we review the construction of the worldsheet action for IIA string theory on $AdS_4 \times \mathbb{CP}^3$ as a super-coset
\begin{equation}
  \frac {OSp(6|4)}{SO(1,3) \times U(3)}
\end{equation}
following \cite{Arutyunov:2008if, Stefanski:2008ik,Zarembo:2009au}
and work it out in component fields up to the sixth order.

This coset sigma model is sufficient for quantizing the string around a background where the string moves in the $\mathbb{CP}^3$ directions. The full Green-Schwarz action for the Type IIA background $AdS_4 \times \mathbb{CP}^3$ was constructed in \cite{Gomis:2008jt} and its gauge fixing was studied in \cite{Grassi:2009yj}, but for our purposes the coset approach will be sufficient.

Coordinates $\xi$ on the coset supermanifold are defined by a coset representative $g(\xi)$, up to gauge transformations $g(\xi) \to g(\xi) h(\xi)$, where $h(\xi)$ is an element of the $SO(1,3) \times U(3)$ group while the global $OSp(6|4)$ transformations act by multiplication from the left $g(\xi) \to g' g(\xi)$.  The building blocks of the worldsheet action are the left-invariant $OSp(6|4)$ currents, defined by
\begin{equation}
  \label{eq:osp_currents}
  j_\mu(\sigma) = g^{-1}(\xi(\sigma)) \partial_\mu g(\xi(\sigma)).
\end{equation}  The Lie superalgebra $osp(6|4)$ admits a $\mathbb{Z}_4$ grading (see \cite{Arutyunov:2008if, Stefanski:2008ik} for more details) under which the left-invariant currents
decompose as $j_\mu = j_\mu^{(0)} + j_\mu^{(1)} + j_\mu^{(2)} + j_\mu^{(3)}$, where the superscript indicates the grading.  The components $j^{(0)}$ and $j^{(2)}$ belong to the even part of the superalgebra while the components $j^{(1)}$ and $j^{(3)}$ belong to the odd part of the superalgebra.  The current $j_\mu^{(0)}$ takes values in the Lie algebra $so(1,3) \times u(3)$ of the denominator of the coset.

Using the $\mathbb{Z}_4$ grading we can easily see that the currents $j_\mu^i$ with $i=1,2,3$ transform homogeneously under the gauge transformation $g(\xi) \to g(\xi) h(\xi)$, i.e. $j_\mu^i \to h^{-1} j_\mu^i h$.  Then, the Lagrangian
\begin{equation}
  \label{eq:sigma_model_action}
  \mathcal{L} = \frac {\sqrt{2 \lambda}} 4 \Str\left(\sqrt{-h} h^{\mu \nu} j_\mu^{(2)} j_\nu^{(2)} + \epsilon^{\mu \nu} j_\mu^{(1)} j_\nu^{(3)}\right)
\end{equation} is invariant under both the gauge symmetry $g(\xi) \to g(\xi) h(\xi)$ and under the global symmetry $g(\xi) \to g' g(\xi)$.
\footnote{In the formula above one should not confuse $h$, the determinant of the worldsheet metric $h_{\mu \nu}$, with the element
of the gauge group $h(\xi)$.}

We will be interested in the tree-level scattering of bosons, so let us truncate the theory to its bosonic sector.  The algebra in the denominator of the coset model is generated by $so(1,3) \times u(3)$ generators $K_i, T_i$ and $R_a^b, R^a, R_a, R,$ where $i=1,2,3$ and $a, b=1,2.$ \footnote{Here we use the same notations and conventions   as in~\cite{Zarembo:2009au}.}  The remaining generators $L_i, D$ complete the algebra $so(1,3)$ to $so(2,3)$ and $B^a, B_a, J, M$ complete the algebra $u(3)$ to $su(4).$ We present the commutation relations for all these generators in Appendix~\ref{sec:commutation_relations}.

In terms of these generators, the coset representative is given by
\begin{equation}
  \label{eq:coset_representative}
  g = \exp \left(t D + \frac i 2 \varphi J\right) \exp \left(\frac 1 {\sqrt{2}} X^a B_a + \frac 1 {\sqrt{2}} \bar{X}_a B^a + \frac 1 2 Z M + Y^i L_i\right),
\end{equation} where the numerical factors have been chosen such that the fields will have the canonical normalization in the action.  This representative of the coset is adapted to the case where the motion of the center of mass of the string is along the light-like geodesic $t = \varphi$.  Upon gauge fixing the fields $t$ and $\varphi$ will be eliminated and one will be left with eight transverse bosonic degrees of freedom, given by the fields $X^a$, $\bar{X}_a$, $Z$ and $Y^i$.

Using the commutation relations in Appendix~\ref{sec:commutation_relations} in the eq.~\eqref{eq:sigma_model_action} and expanding the action in powers of the transverse fields $X^a$, $\bar{X}_a$, $Z$ and $Y^i$, we obtain the Lagrangian ${\mathcal L}$
\be
{\mathcal L}=\sum_{i} {\mathcal L}^{(i)},
\ee
where
{\allowdisplaybreaks
\begin{subequations}\label{eq:lagrangian_expansion}
\begin{align}
  \mathcal{L}^{(0)} = & \frac{\partial_\mu \varphi^2} 2-\frac{\partial_\mu t^2} 2,\\
  \mathcal{L}^{(2)} = & \partial_\mu X \cdot \partial^\mu \bar{X} + \frac 1 2 (\partial_\mu Y \cdot \partial^\mu Y) - \frac 1 2 Y \cdot Y (\partial_\mu t)^2 + \frac{\partial_\mu Z^2}{2}-\frac{1}{4} \left(2 Z^2+X \cdot \bar{X}\right) \partial_\mu \varphi^2,\\
  \mathcal{L}^{(3)} = & -\frac 1  2  i Z \left(X \cdot \partial_\mu \bar{X}-\bar{X} \cdot \partial_\mu X\right) \partial^\mu \varphi,\\
  \mathcal{L}^{(4)} = & \begin{aligned}[t]& \frac{1}{6} \left(X \cdot \partial_\mu \bar{X}\right)^2-\frac{1}{6}
   \bar{X} \cdot \partial_\mu X X \cdot \partial^\mu \bar{X}+\frac{1}{6} \left(\bar{X} \cdot \partial_\mu X\right)^2- \frac 1 {12} X \cdot \bar{X} \partial_\mu Z^2+ \\+ & \frac{1}{48} \left(8 Z^2+X \cdot \bar{X}\right) \left(Z^2+2 X \cdot \bar{X}\right) \partial_\mu \varphi^2-\frac{1}{12} \left(Z^2+2 X \cdot \bar{X}\right) \partial_\mu X \cdot \partial^\mu \bar{X}+\\ +& \frac 1 6 (Y \cdot Y) (\partial_\mu Y \cdot \partial^\mu Y) - \frac 1 6 (Y \cdot Y)^2 (\partial_\mu t)^2 - \frac 1 6 (Y \cdot \partial_\mu Y)^2 +\\
+ & \frac{1}{12} Z  \partial_\mu Z \partial^\mu (X \cdot \bar{X} ),
\end{aligned}\\
  \mathcal{L}^{(5)} = & \frac 1 8 i Z \left(Z^2+2 X \cdot \bar{X}\right) \left(X \cdot \partial_\mu \bar{X}-\bar{X} \cdot \partial_\mu X\right) \partial^\mu \varphi,\\
  \mathcal{L}^{(6)} = & \begin{aligned}[t] & \frac 1 {1440}  \left(Z^2+2 X \cdot \bar{X}\right)^2 \left(4 \partial_\mu X \cdot \partial^\mu \bar{X}-\left(32 Z^2+X \cdot \bar{X}\right) \partial_\mu \varphi^2\right)+ \\
+ & \frac 1 {360} \left(Z^2+2 X \cdot \bar{X}\right) \bigl(-8 \left(X \cdot \partial_\mu \bar{X}\right)^2+14 \partial_\mu X \cdot \bar{X}  X \cdot \partial_\mu \bar{X}- \\&\hphantom{\frac 1 {360} \left(Z^2+2 X \cdot \bar{X}\right) \bigl(}-8  \left(\bar{X} \cdot \partial_\mu X\right)^2+ X \cdot \bar{X} \partial_\mu Z^2-Z  \partial_\mu Z \partial^\mu (X \cdot \bar{X}) \bigr) +\\& + \frac 1 {45} (Y \cdot Y)^2 (\partial_\mu Y \cdot \partial^\mu Y) - \frac 1 {45} (Y \cdot Y)^3 (\partial_\mu t)^2 - \frac 1 {45} (Y \cdot Y) (Y \cdot \partial_\mu Y)^2.
\end{aligned}
\end{align}
\end{subequations}}

The Lagrangian above can be written in the following generic form
\begin{equation}
  \label{eq:lagrangian_metric_components}
  \mathcal{L} =
  \begin{aligned}[t]
    \frac 1 2 \Big(& G_{a b} \partial_\mu X^a \partial^\mu X^b + G^{a b} \partial_\mu \bar{X}_a \partial^\mu \bar{X}_b + G^a_{\hphantom{a} b} \partial_\mu \bar{X}_a \partial^\mu X^b + G_b^{\hphantom{b} a} \partial_\mu \bar{X}_a \partial^\mu X^b +\\
& + 2 G_{a Z} \partial_\mu X^a \partial^\mu Z + 2 G^a_{\hphantom{a} Z} \partial_\mu \bar{X}_a \partial^\mu Z + G_{i j} \partial_\mu Y^i \partial^\mu Y^j + G_{Z Z} \partial_\mu Z \partial^\mu Z +\\
& + 2 G_{\varphi a} \partial_\mu \varphi \partial^\mu X^a + 2 G_\varphi^{\hphantom{\varphi} a} \partial_\mu \varphi \partial^\mu \bar{X}_a + 2 G_{\varphi Z} \partial_\mu \varphi \partial^\mu Z +\\
& + G_{\varphi \varphi} \partial_\mu \varphi \partial^\mu \varphi - G_{t t} \partial_\mu t \partial^\mu t\Big),
  \end{aligned}
\end{equation} from which we can extract the components of the metric tensor up to the required order
{\allowdisplaybreaks
\begin{subequations}\label{eq:metric_components}
\begin{align}
  G_{a b} = & (G^{a b})^* = \left(\frac 1 3 - \frac 2 {45} Z^2 - \frac 4 {45} X \cdot \bar{X} + \dotso \right) \bar{X}_a \bar{X}_b,\\
  G^a_{\hphantom{a} b} = & G_b^{\hphantom{b} a} = \begin{aligned}[t]&\left(1 - \frac {Z^2}{12} + \frac {Z^4}{360} - \frac {X \cdot \bar{X}} 6 + \frac 1 {90} Z^2 X \cdot \bar{X} + \frac 1 {90} (X \cdot \bar{X})^2 + \dotso\right) \delta^a_b +\\&+ \left(-\frac 1 6 + \frac {7 Z^2}{180} + \frac {7 X \cdot \bar{X}}{90} + \dotso\right) X^a \bar{X}_b,\end{aligned}\\
  G_{i j} = & \delta_{i j} + \left(\frac 1 3 + \frac 2 {45} Y \cdot Y + \dotso\right) (Y \cdot Y \delta_{i j} - Y_i Y_j),\\
  G_{\varphi a} = & (G_\varphi^{\hphantom{\varphi} a})^* = 
i \bar{X}_a \Bigl(\frac Z 2 - \frac {Z^3} 8 - \frac {Z X \cdot \bar{X}} 4 +
 \dotso\Bigr),\\
  G_{Z a} = & (G_Z^{\hphantom{Z} a})^* = \left(\frac Z {12} - \frac {Z^3} {360} - \frac {Z X \cdot \bar{X}}{180} + \dotso\right) \bar{X}_a,\\
  G_{Z Z} = & 1 - \frac {X \cdot \bar{X}} 6 + \frac 1 {180} Z^2 X \cdot \bar{X} + \frac 1 {90} (X \cdot \bar{X})^2 + \dotso,\\
  G_{\varphi \varphi} = & \begin{aligned}[t] &1 - Z^2 - \frac {X \cdot \bar{X}} 2  + \frac {Z^4} 3 + \frac {17}{24} Z^2 X \cdot \bar{X} + \frac 1 {12} (X \cdot \bar{X})^2 -\\&- \frac 2{45} Z^6 - \frac {43}{240} Z^4 X \cdot \bar{X} - \frac {11}{60} Z^2 (X \cdot \bar{X})^2  - \frac 1 {180} (X \cdot \bar{X})^3 + \dotso, \end{aligned}\\
  G_{t t} = & 1 + Y \cdot Y + \frac 1 3 (Y \cdot Y)^2 + \frac 2 {45} (Y \cdot Y)^3 + \dotso.
\end{align}
\end{subequations}}

\section{Gauge fixing}
\label{sec:gauge_fixing}

In this section we gauge-fix the action~(\ref{eq:lagrangian_metric_components}) derived in the previous section and work it out in components up to the sixth order.

Let us start by reviewing following \cite{Zarembo:2009au} the light-cone gauge fixing for the
bosonic string in the background
\begin{equation}
 ds^2=-G_{tt}dt^2+G_{\varphi\varphi}d\varphi^2+2G_{\varphi A}d\varphi dx^A+G_{A B}d x^A d x^B,
\end{equation}
where $G_{tt}$, $G_{\varphi \varphi }$, $G_{\varphi A}$ and $G_{A B}$
are functions of $x^A$.

We will consider the $a$-gauge, a one-parameter family of interpolating gauges introduced in~\cite{Arutyunov:2006gs}.  The temporal and light-cone gauges correspond to $a=0$ and $a=1/2$ respectively.

Gauge-fixing can be done in several different but equivalent ways \cite{Arutyunov:2006gs,Arutyunov:2009ga,Kruczenski:2004cn,Zarembo:2009au}. A very convenient approach is to use a trick where one $T$-dualizes first in a direction $X^- = a t - (1-a) \varphi$, thereby replacing the coordinate $X^-$ with its $T$-dual coordinate $\tilde{\varphi}$. Then, by integrating out the worldsheet metric we obtain a Nambu-Goto action.  We can then use the gauge-fixing conditions
\begin{equation}
  \label{eq:gauge-fixing-conditions}
  X^+ \equiv (1-a) t + a \varphi = \tau, \quad \tilde{\varphi} = \sigma,
\end{equation} where $(\tau, \sigma)$ are worldsheet coordinates, we get the following gauge-fixed action (this formula was used in~\cite{Zarembo:2009au} as the starting point for the worldsheet scattering computations):

\begin{eqnarray}\label{ng}
 \mathcal{L}_{\rm NG}&=&
 -\frac{1}
 {\left(1-a\right)^2G_{\varphi\varphi}-a^2G_{tt}}
 \left\{\vphantom{\frac{aG_{tt}}{1-a}}
 \left[\vphantom{+\left(aG_{tt}G_{\varphi A}\partial _1 X^A
 -\tilde{G}_{A B}\partial_1 X^i \partial_1 X^B \right)^2}
 \left(
 G_{tt}G_{\varphi \varphi }+2aG_{tt} G_{\varphi A}\partial_0 X^A-
 \tilde{G}_{A B}\partial_0 X^A \partial_0 X^B\right)
 \right.\right. \nonumber \\
 && \left.\left.\times
 \left(1+\tilde{G}_{A B} \partial_1 X^A \partial_1 X^B\right)
 +\left(aG_{tt}G_{\varphi A} \partial_1 X^A
 -\tilde{G}_{A B} \partial_0 X^A \partial_1 X^B\right)^2
 \right]^{1/2}
 \right.\nonumber \\ &&\left.
 -\frac{aG_{tt}}{1-a}-\left(1-a\right)G_{\varphi A}\partial_0 X^A
 \right\},
\end{eqnarray}
where the metric $\tilde{G}_{A B}$ is
\begin{equation}\label{}
 \tilde{G}_{A B}=\left[\left(1-a\right)^2G_{\varphi\varphi}-a^2G_{tt}
 \right]G_{A B}-\left(1-a\right)^2G_{\varphi A}G_{\varphi B}.
\end{equation}

After the gauge fixing the two bosonic sectors $AdS_4$ and $\mathbb{CP}^3$ become coupled.  This is similar to studies of classical strings where
two otherwise separate bosonic sectors are coupled through Virasoro constraints.

Expanding this Lagrangian to sixth order we get
{\allowdisplaybreaks
\begin{equation}\begin{aligned}
 \mathcal{L} &=
 \frac{G_{A B}}{2}\,\partial_\mu X^A \partial^\mu X^B - \frac{G_{tt}}{2} + \frac{G_{\varphi \varphi }}{2} + G_{\varphi A}\partial_0 X^A\\
 & +\frac{1}{4}\left(1 - G_{tt}G_{\varphi \varphi}\right) G_{A B} \left(\partial_0 X^A \partial_0 X^B + \partial_1 X^A \partial_1 X^B\right) + \frac{1}{4}\left(G_{tt} - 1\right)^2 - \frac{1}{4}\left(G_{\varphi \varphi} - 1\right)^2\\
 & -\frac{1-2a}{8}\left({G_{tt}} - {G_{\varphi \varphi}}\right)^2 - \frac{1-2a}{8}\left(G_{A B}\partial_\mu X^A \partial^\mu X^B\right)^2 + \frac{1-2a}{4}\left(G_{A B}\partial_\mu X^A \partial_\nu X^B\right)^2\\
 & -\frac{1}{2}(a (G_{tt}-1)+(1-2 a)(G_{\varphi \varphi}-1)) G_{\varphi A}X^A-\\
 & -\frac{a}{2} G_{A B} G_{\varphi C} (\partial_0 X^A \partial_0 X^B \partial_0 X^C+ \partial_1 X^A \partial_1 X^B \partial_1 X^C - 2 \partial_0 X^A \partial_1 X^B \partial_1 X^C)\\
 &-\frac{(1-2a)^2}{32}\Bigl[(G_{tt}-G_{\varphi \varphi})^3 + \bigl((G_{tt}-G_{\varphi \varphi}) - 2 G_{A B}\partial_{\mu} X^A \partial^{\mu} X^B\bigr)\times\\
 &\hphantom{-\frac{(1-2a)^2}{32}\Bigl[(G_{tt}-G_{\varphi \varphi})^3} \bigl(2 (G_{A B}\partial_{\mu} X^A \partial_{\nu} X^B)^2 -(G_{A B}\partial_{\mu} X^A \partial^{\mu} X^B)^2 \bigl)\Bigr] + \\
 &+\frac{1-2a}{8}(G_{tt}+G_{\varphi \varphi}-2) \Bigl[(G_{tt}-G_{\varphi \varphi})^2 - (G_{A B}\partial_{0} X^A \partial_{0} X^B)^2+(G_{A B}\partial_{1} X^A \partial_{1} X^B)^2 \Bigr]+\\
 &+\frac{1}{32} \Bigl[(G_{\varphi \varphi} - G_{tt}) (2 (G_{t t} - 1)^2 + 2 (G_{\varphi \varphi} - 1)^2 + 3 (2 - G_{t t} - G_{\varphi \varphi})^2)-\\
 &\hphantom{+\frac{1}{32} \Bigl[}-16 G_{\varphi A} G_{\varphi B} \partial_\mu X^A \partial^\mu X^B+\\
 &\hphantom{+\frac{1}{32} \Bigl[}+ 2 (-2 + G_{\varphi \varphi} + G_{tt})^2 (3 G_{A B}\partial_{0} X^A \partial_{0} X^B+G_{A B}\partial_{1} X^A \partial_{1} X^B)+\\
 &\hphantom{+\frac{1}{32} \Bigl[}+ (G_{\varphi \varphi} - G_{tt})(2 (G_{A B}\partial_{\mu} X^A \partial_{\nu} X^B)^2 -(G_{A B}\partial_{\mu} X^A \partial^{\mu} X^B)^2 )\Bigr]+\dotso. \label{eq:L_sixth-order}
\end{aligned}\end{equation}}%

Plugging in the components of the metric~\eqref{eq:metric_components} worked out in the previous section, we can easily obtain the gauge fixed Lagrangian up to sixth order.  In order to transcribe~\eqref{eq:metric_components} into this formula we use $G_{A B}=\lbrace G_{ab}, G^{ab}, G^a_{\hphantom{a} b}, G_b^{\hphantom{b} a}, G_{i j}, G_{ZZ}, G_{Za}, G_Z^a\rbrace$, $G_{\varphi A}=\lbrace G_{\varphi a}, G_{\varphi}^a, G_{\varphi   Z}\rbrace$ and $X^A = \lbrace X^a, \bar{X}_a, Y^i, Z\rbrace$.

Let us present the fourth order gauge-fixed Lagrangian for the fields $X$, $Y$ and $Z$.  Because the formulas are lengthy, we find it convenient to separate this Lagrangian into pieces, according to the field content.
\begin{equation}
  \label{eq:L4_gf_XXXX}
  \begin{aligned}
    \mathcal{L}_{XXXX}^{(4)} &= \frac {-5 + 6 a}{96} (X \cdot \bar{X})^2 + \frac 1 6 (X \cdot \partial_\mu \bar{X})^2 + \frac 1 6 (\partial_\mu X \cdot \bar{X})^2 - \frac 1 6 (X \cdot \partial_\mu \bar{X}) (\partial^\mu X \cdot \bar{X}) +\\&+ \frac 1 {12} X \cdot \bar{X} (\partial_0 X \cdot \partial_0 \bar{X} + 5 \partial_1 X \cdot \partial_1 \bar{X}) +\\& + \frac {1 - 2 a} 2 \left[- (\partial_\mu X \cdot \partial^\mu \bar{X})^2 + (\partial_\mu X \cdot \partial_\nu \bar{X})^2 + (\partial_\mu X \cdot \partial_\nu \bar{X}) (\partial^\nu X \cdot \partial^\mu \bar{X})\right],
  \end{aligned}
\end{equation}
\begin{equation}
  \label{eq:L4_gf_YYYY}
  \begin{aligned}
    \mathcal{L}_{YYYY}^{(4)} &= \frac {-1 + 6 a}{24} (Y \cdot Y)^2 - \frac 1 6 (Y \cdot \partial_\mu Y)^2 - \frac 1 {12} Y \cdot Y (\partial_0 Y \cdot \partial_0 Y + 5 \partial_1 Y \cdot \partial_1 Y) +\\& + \frac {1 - 2 a} 8 \left[- (\partial_\mu Y \cdot \partial^\mu Y)^2 + 2 (\partial_\mu Y \cdot \partial_\nu Y)^2\right],
  \end{aligned}
\end{equation}
\begin{equation}
  \label{eq:L4_gf_ZZZZ}
    \mathcal{L}_{ZZZZ}^{(4)} = \frac {-5 + 6 a}{24} Z^4 + \frac 1 4 Z^2 \left((\partial_0 Z)^2 + (\partial_1 Z)^2\right) + \frac {1 - 2 a} 8 (\partial_\mu Z \partial^\mu Z)^2,
\end{equation}
\begin{equation}
  \label{eq:L4_gf_XXYY}
  \begin{aligned}
    \mathcal{L}_{XXYY}^{(4)} &= - \frac {1 - 2 a} 8 X \cdot \bar{X} Y \cdot Y - \frac 1 2 Y \cdot Y (\partial_0 X \cdot \partial_0 \bar{X} + \partial_1 X \cdot \partial_1 \bar{X}) + \\&+ \frac 1 8 X \cdot \bar{X} (\partial_0 Y \cdot \partial_0 Y + \partial_1 Y \cdot \partial_1 Y) + \\&+ \frac {1 - 2 a} 2 \left[- (\partial_\mu X \cdot \partial^\mu \bar{X}) (\partial_\nu Y \cdot \partial^\nu Y) + 2 (\partial_\mu X \cdot \partial_\nu \bar{X}) (\partial^\mu Y \cdot \partial^\nu Y)\right],
  \end{aligned}
\end{equation}
\begin{equation}
  \label{eq:L4_gf_XXZZ}
  \begin{aligned}
    \mathcal{L}_{XXZZ}^{(4)} &= - \frac {1 - 6 a} {48} Z^2 X \cdot \bar{X} - \frac 1 {12} Z \partial_\mu Z (\partial^\mu X \cdot \bar{X} + X \cdot \partial^\mu \bar{X}) +\\&+ \frac 1 {12} Z^2 (5 \partial_0 X \cdot \partial_0 \bar{X} + 7 \partial_1 X \cdot \partial_1 \bar{X}) + \frac 1 {24} X \cdot \bar{X} \left((\partial_0 Z)^2 + 5 (\partial_1 Z)^2\right) +\\&+ \frac {1 - 2 a} 2 \left[- (\partial_\mu Z \partial^\mu Z) (\partial_\nu X \cdot \partial^\nu \bar{X}) + 2 (\partial_\mu Z \partial_\nu Z) (\partial^\mu X \cdot \partial^\nu \bar{X})\right],
  \end{aligned}
\end{equation}
\begin{equation}
  \label{eq:L4_gf_YYZZ}
  \begin{aligned}
    \mathcal{L}_{YYZZ}^{(4)} &= - \frac {1 - 2 a} 4 Y \cdot Y Z^2 + \frac 1 4Z^2 (\partial_0 Y \cdot \partial_0 Y + \partial_1 Y \cdot \partial_1 Y) - \frac 1 4 Y \cdot Y \left((\partial_0 Z)^2 + (\partial_1 Z)^2\right) +\\&+ \frac {1 - 2 a} 4 \left[- (\partial_\mu Y \cdot \partial^\mu Y) (\partial_\nu Z \partial^\nu Z) + 2 (\partial_\mu Y \cdot \partial_\nu Y) (\partial^\mu Z \partial^\nu Z)\right].
  \end{aligned}
\end{equation}

The full gauge-fixed fourth order Lagrangian is
\begin{equation}
  \label{eq:L4_gf}
  \mathcal{L}_{\text{gf}}^{(4)} = \mathcal{L}_{XXXX}^{(4)} + \mathcal{L}_{YYYY}^{(4)} + \mathcal{L}_{ZZZZ}^{(4)} + \mathcal{L}_{XXYY}^{(4)} + \mathcal{L}_{XXZZ}^{(4)} + \mathcal{L}_{YYZZ}^{(4)}.
\end{equation}

For reasons of space we choose not to present the expression for the order five and six Lagrangian.  The higher order terms can be found by using eqs.~\eqref{eq:metric_components} in the eq.~\eqref{eq:L_sixth-order}.  The fourth order term~\eqref{eq:L4_gf_XXXX}
with four $X$ fields was
derived by Zarembo in~\cite{Zarembo:2009au}.

\section{Scattering amplitudes}
\label{sec:scattering_amplitudes}

In this section we will use the Feynman rules which follow from the Lagrangian derived in the previous section to compute four-, five- and six-point scattering amplitudes at tree level both analytically and numerically.

Except for non-vanishing four-point amplitudes, we adopt the convention that all momenta are incoming and we parameterize
the on-shell energy and momenta for the $i$-th particle as follows
\begin{equation}
  \e_i = \frac {m_i} 2 \left(a_i + \frac 1 {a_i}\right), \quad
  p_i = \frac {m_i} 2 \left(a_i - \frac 1 {a_i}\right).
\end{equation}

We will consider scattering amplitudes of $X,$ $Y$ and $Z$ fields.  As argued by Zarembo, the $Z$ are not expected to survive as physical states once $\alpha'$ corrections are taken into account.  However, at tree level, which is the case we are interested in, the $Z$ particles can appear as asymptotic states and it makes sense to talk about their $S$-matrix. (There is an interaction term $(X^a \overset{\leftrightarrow}{\partial}_0 \bar{X}_a) Z$ in the Lagrangian but one can check that this interaction vanishes on-shell so the field $Z$ is stable at tree level.  Therefore, at tree level, it makes sense to include the $Z$ states as asymptotic states.)
Similar considerations apply to $Y$ fields.

In the following sections we compute all bosonic four-, five- and six-point tree-level amplitudes.  In writing these amplitudes, we leave out the momentum conservation delta function and the external leg factors $\tfrac 1 {\sqrt{2 \epsilon}}$.  We compute only the connected part of the amplitude.
In several cases the amplitude can be simplified by using symmetric polynomials
(see Appendix~\ref{sec:toy_model} for a toy model example).

\subsection{Four-point amplitudes}

\begin{figure}
  \centering
  \includegraphics{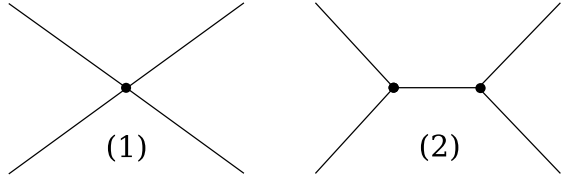}
  \caption{The four-point master topologies.  The first topology is the contact term, while the second topology generates the $s$-, $t$- and $u$-channels by permutations of the external legs.  Depending on the external states, not all of the diagrams obtained by permutations of the external states will give a non-vanishing contribution.}.
  \label{fig:4point}
\end{figure}

The scattering amplitudes with four $X$ fields have been computed in~\cite{Zarembo:2009au},
so we will proceed with the remaining cases.

\subsubsection*{$\bullet$ $X^2 Z^2$}

In the case of $X^a \bar{X}_b ZZ$ we denote the momenta by $p_i$, with $i = 1, \dotsc, 4$, where $p_1$ and $p_2$ are the momenta of the particles $X$ and $\bar{X}$, with $\varepsilon_1^2 - p_1^2 = \varepsilon_2^2 - p_2^2 = 1/4$ and $p_3$ and $p_4$ are the momenta of the particles $Z$, with $\varepsilon_3^2 - p_3^2 = \varepsilon_4^2 - p_4^2 = 1$.  The corresponding light-cone momenta $a_i$ must also satisfy momentum conservation
\be\ba
  a_1 + a_2 + 2 a_3 + 2 a_4 &= 0,\\
  \frac 1 {a_1} + \frac 1 {a_2} + \frac 2 {a_3} + \frac 2 {a_4} &= 0.
\ea\ee

The expression for the $X^a \bar{X}_b  ZZ$ scattering amplitude obtained from the Feynman rules computation is fairly complicated, but can be simplified by using momentum conservation and on-shell conditions.  One way of writing this amplitude is
\begin{equation}
  \begin{aligned}
    i \delta_b^a\Bigg\lbrace&\left(-\frac {1 - (a_3+a_4)^2 + a_3^2 a_4^2}{4 a_3 a_4}\right)_c +\\
    & \left(-\frac {(a_1 - a_2)(a_3 - a_4)}{16 a_3 a_4} + \frac {1 - (a_3+a_4)^2 + a_3^2 a_4^2}{8 a_3 a_4}\right)_t +\\
    & \left(\frac {(a_1 - a_2)(a_3 - a_4)}{16 a_3 a_4} + \frac {1 - (a_3+a_4)^2 + a_3^2 a_4^2}{8 a_3 a_4}\right)_u\Bigg\rbrace=0,
  \end{aligned}
\end{equation} where the $c$, $t$ and $u$ subscripts mark the contributions of the contact term and of the $t$- and $u$-channels respectively.  It is easy to see that all these contributions from different Feynman diagrams cancel when added together.

\subsubsection*{$\bullet$ $Z^4$}

The scattering amplitude for the $ZZ \to ZZ$ process is given by
\begin{equation}
  S = 2 i \left((p_1^2+p_2^2) + (1 - 2 a) (p_1 \varepsilon_2-p_2 \varepsilon_1)^2\right),
\end{equation} where $(\varepsilon_1, p_1)$ and $(\varepsilon_2, p_2)$ are the on-shell incoming momenta ($\varepsilon_i^2 - p_i^2 = 1$).  Note that the piece multiplying $(1 - 2 a)$ is Lorentz invariant, just like the interaction term in eq.~\eqref{eq:L4_gf_ZZZZ}.

\subsubsection*{$\bullet$  $Y^4$}

\begin{figure}
  \centering
  \includegraphics{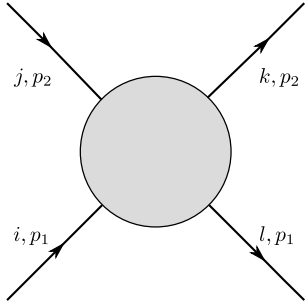}
  \caption{The scattering of four $Y$ fields.  The indices $i$, $j$, $k$ and $l$ run from $1$ to $3$.}
  \label{fig:4Y}
\end{figure}

For the $Y^i Y^j \to Y^k Y^l$ scattering process in Fig.~\ref{fig:4Y} the amplitude becomes
\be
S=4 i p_1 p_2 ( \delta^{ij}\delta^{kl}-\delta^{ik}\delta^{jl}) - 2 i \left( p_1^2+p_2^2-(1-2a) (p_1 \e_2-p_2 \e_1)^2 \right) \delta^{il}\delta^{jk},
\ee
where the momenta of the $k$ and $l$ fields are considered to be outgoing.

\subsubsection*{$\bullet$ Remaining four-point amplitudes}

We have also numerically checked that the scattering processes $Y^i Y^j ZZ$ and $X^a \bar{X}_b Y^i Y^j$ vanish at several kinematic points.

\subsection{Five-point amplitudes}

\FIGURE{\includegraphics{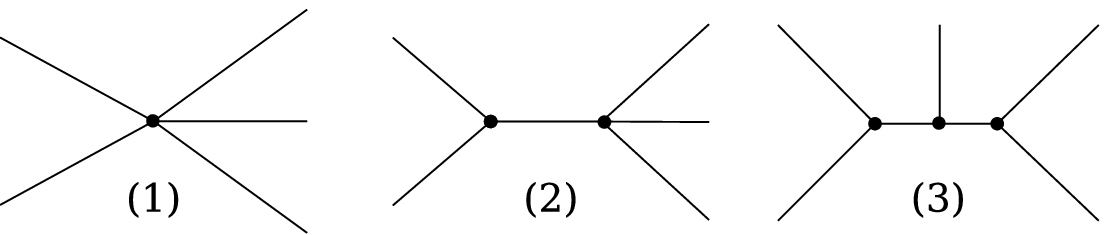}%
  \caption{The five-point master topologies.  All the five-point Feynman diagrams can be obtained from these master topologies by permutation of external legs.  The first topology yields a single diagram, the second topology yields $10$ diagrams and the third topology yields $15$ diagrams.}%
  \label{fig:5point}}

\subsubsection*{$\bullet$ $X^2 Z^3$}

In the $X^a \bar{X}_b  ZZZ$ scattering process we have three master topologies as illustrated in Fig.~\ref{fig:5point}.  It is possible to obtain analytic results for all three classes of diagrams in terms of symmetric polynomials.  To this end we parameterize the momentum of the $Z$ particles by $a_i$ with $i=1,2,3$, and the momentum of $X,\bar{X}$ by $a_4$ and $a_5$ respectively.  Conservation of momentum then yields the constraints
\be\ba
2a_1+2a_2+2a_3+a_4+a_5 &=0,\\
\frac 2 {a_1} + \frac 2 {a_2} + \frac 2 {a_3} + \frac 1 {a_4} +\frac 1 {a_5} &= 0.
\ea\ee
The contributions of the three master topologies to the $S$-matrix are
\be\ba
(S_1)^a_b&= i\, (A+(1-2a)B) \, \delta^a_b,\\
(S_2)^a_b&=- i\, (A+C+(1-2a)B) \, \delta^a_b,\\
(S_3)^a_b&= i\, C \, \delta^a_b,
\ea\ee
where
\be\ba
A&=\frac{s_1^2+6 s_1 s_3-s_2 \left(s_2+6\right)}{8 s_3}\left(1-\frac{s_3}{s_1 s_2}\right)^{1/2},\\
B&=\frac{-s_1^2+3 s_1 s_3+s_2 \left(s_2-3\right)}{8 s_3}\left(1-\frac{s_3}{s_1 s_2}\right)^{1/2},\\
C&=\frac 1 {4 s_3 \left(s_3 s_1^3+2 s_2^2 s_1^2+s_2^3\right)} \left(1-\frac{s_3}{s_1 s_2}\right)^{-1/2} \Big\lbrace  \left(7 s_2^2+6 s_2-3 s_3^2+3\right) s_1 s_2 s_3+2 s_2^2 \left(s_2^2-3 s_3^2\right)\\
& \quad +\left(3 s_2^2-2 s_3^2\right) s_1^4+ \left(2 s_2-7\right)s_2 s_3 s_1^3-\left(3 s_2^4+2 s_2^3-3 \left(s_3^2-1\right) s_2^2+6 s_3^2 s_2-6 s_3^2\right) s_1^2 \Big\rbrace,
\ea\ee
and $s_i$ symmetric polynomials with respect to $a_1,a_2,a_3$ variables.  We can easily see that the amplitude vanishes as expected, that is
\be
(S_1)^a_b+(S_2)^a_b+(S_3)^a_b=0.
\ee

\subsubsection*{$\bullet$ $X^4 Z$}

Another 5-point scattering process is $X^a \bar{X}_b X^c\bar{X}_d Z$. If we parameterize $Z$ with $a_5$ and $X^a,\bar{X}_b,X^c,\bar{X}_d$ with $a_1,a_2,a_3,a_4$ the energy-momentum conservation imposes the constraints
\be\label{constraint}\ba
a_1+a_2+a_3+a_4+2a_5 &=0,\\
\frac 1 {a_1} + \frac 1 {a_2} + \frac 1 {a_3} + \frac 1 {a_4} +\frac 2 {a_5} &= 0.
\ea\ee

Although we have analytic expressions for all three classes of master topologies as they are shown in Fig.~\ref{fig:5point} we choose to explicitly show only the result for the contact term interaction.  Then the sum of the second and third of the diagrams in Fig.~\ref{fig:5point} precisely cancels the contact term yielding zero as expected.  The contact term contribution to the scattering is
\begin{multline}\label{S1}
(S_1)^{ac}_{bd}=i \frac {a} {32} \Bigg\lbrace \frac{ \left(\left(r_2-1\right) r_3+r_1 \left(r_4-1\right)\right) \left(r_2+r_4\right)}{ r_2 r_4} \delta^a_b \delta^c_d +\\ \frac{ \left(t_1 \left(t_3-1\right)-t_2 \left(t_4-1\right)\right) \left(t_3+t_4\right)}{ t_3 t_4} \delta^a_d \delta^c_b \Bigg\rbrace,
\end{multline} where
\begin{alignat*}{4}
r_1 &= a_1-a_2, &\quad r_2 &= a_1 a_2, &\quad r_3 &= a_3-a_4, &\quad r_4 &= a_3 a_4,\\
t_1 &= a_1-a_4, &\quad t_4 &= a_1 a_4, &\quad t_2 &= a_2-a_3, &\quad t_3 &= a_2 a_3.
\end{alignat*}

\subsubsection*{$\bullet$ Remaining five-point amplitudes}

We have also numerically checked that the scattering processes $Y^i Y^j Y^k Y^l Z$, $Y^i Y^j Z Z Z$ and $X^a \bar{X}_b Y^i Y^j Z$ vanish at several kinematic points.

\subsection{Six-point amplitudes}

\FIGURE{\includegraphics{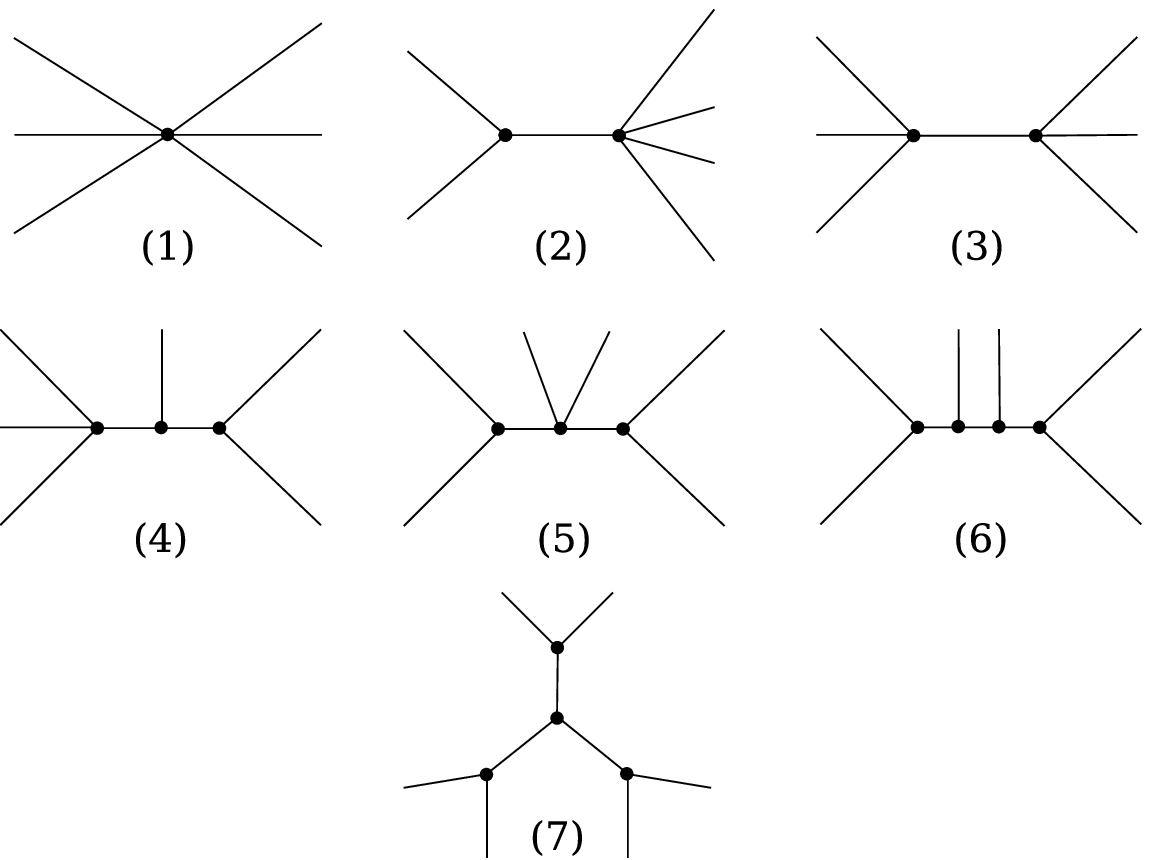}%
  \caption{The seven six-point master topologies.  These topologies contribute $1$, $15$, $10$, $60$, $45$, $90$ and $15$ diagrams respectively.  In total there are $236$ diagrams.  Depending on the external states not all the topologies or all the diagrams belonging to a given topology contribute to the scattering amplitude.}%
  \label{fig:6point}}

\subsubsection*{$\bullet$ $Z^6$ }

The energy-momentum conservation gives
\be
\sum_{i=1}^6 a_i = \sum_{i=1}^6 \frac{1}{a_i}=0.
\ee
For this scattering process it turns out that only the first and third class of master topologies in Fig.~\ref{fig:6point} contribute and that they precisely cancel each other.  The contact term of this scattering is given by
\be
S_1=-i \frac{\left(s_2+4\right) s_4+\left(4 s_2+7\right) s_6}{4 s_6} -3 i (1-2a) \left( s_2+\frac{ s_4}{s_6}+10\right) - i \frac{(1-2a)^2 \left(9 s_3^2-17 s_2 s_4+81 s_6\right)}{4 s_6}.
\ee

We have analytically checked that the sum of all the diagrams vanishes for this process.

\subsubsection*{$\bullet$ $X^2 Z^4$}

For the $X^a \bar{X}_b ZZZZ$ scattering process we parameterize the momenta of $X,\bar{X}$ with $a_1,a_2$ and the momenta of the our $Z$'s with $a_3,a_4,a_5,a_6$.  Then the energy-momentum conservation gives the constraints
\be\ba
a_1+a_2+2a_3+2a_4+2a_5+2a_6 &=0,\\
\frac 1 {a_1} + \frac 1 {a_2} + \frac 2 {a_3} + \frac 2 {a_4} +\frac 2 {a_5} +\frac 2 {a_6} &= 0.
\ea\ee
The contact term that is canceled by the Feynman diagrams (2) through (7) in Fig.~\ref{fig:6point} can be written in terms of symmetric polynomials in the $a_3,a_4,a_5,a_6$ as
\be
(S_1)^a_b=\frac i {16 s_1 s_3 s_4^2}\left(\frac 1 9  A+ \frac 1 3 B(1-2a)+C(1-2a)^2\right) \delta^a_b,
\ee
where
\be\ba
A&=18 s_1 s_3^3-\left(9  \left(s_2-3\right)s_1 s_2+\left(224 s_1^2+9 \left(s_2+2\right)\right) s_3\right) s_4 s_3\\
&\quad -18 s_1^2 s_4^3+s_1 \left(\left(18 s_1^2+27 s_2+560\right) s_3-9 s_1 s_2\right) s_4^2,\\
B&=s_3 s_4 \left(-2 s_2+3 s_4+9\right) s_1^3+\left(4 s_4 \left(7 s_3^2-3 s_4^2\right)+s_2 \left(12 s_3^2 \left(s_4+1\right)-s_4^2\right)\right) s_1^2\\
& \quad + \left(\left(-2 s_2+9 s_4+3\right) s_3^2+4 s_4 \left(5 s_4-2 s_2 \left(s_2+3 s_4+3\right)\right)\right) s_1 s_3-\left(s_2+12\right) s_3^2 s_4,\\
C&=\left(-10 s_3 s_1^2+s_2^2 s_1+5 s_2 s_3\right) s_4 s_3\\
& \quad +\left(8 s_3 s_1^3-4 s_2^2 s_1^2-3 s_3^2\right) s_3^2+ \left(-3 s_1^3+5 s_2 s_1-16 s_3\right)s_1 s_4^2.
\ea\ee

We have numerically checked at several random kinematic points that the sum of all the diagrams vanishes for this process.

\subsubsection*{$\bullet$ $X^4 Z^2$}

For the $X^a\bar{X}_b X^c\bar{X}_d ZZ$ case we parameterize the momentum of the particles with the order appearing with $a_1,a_2,a_3,a_4,a_5,a_6$.  The conservation of momentum gives the two constraints
\be\ba
a_1+a_2+a_3+a_4+2a_5+2a_6 &=0,\\
\frac 1 {a_1} + \frac 1 {a_2} + \frac 1 {a_3} + \frac 1 {a_4} +\frac 2 {a_5} +\frac 2 {a_6} &= 0.
\ea\ee
Then the contact term is
\be\ba
(S_1)^{ac}_{bd}&=\frac {i}{64 r_2 r_4 r_6}\left(\frac{1}{45} A_r + \frac{1}{3} B_r (1-2a)+ C_r (1-2a)^2 \right) \delta^a_b \delta^c_d\\
& \quad + \frac {i}{64 t_2 t_4 t_6}\left(\frac{1}{45} A_t + \frac{1}{3} B_t (1-2a)+ C_t (1-2a)^2 \right) \delta^a_d \delta^c_b,
\ea\ee
where
\be\ba
A_r&=-98 r_4 r_6 r_1^2-\left(30 r_3 r_6+r_4 \left(r_5 \left(15-4 r_6\right)-2 r_3 r_6\right)\right) r_1-15 r_4 r_6 \left(3 r_4+3 r_6-10\right)\\
&\quad -r_2 \left(98 r_6 r_3^2+\left(r_5 \left(15-4 r_6\right)-2 r_1 r_6\right) r_3+r_4^2 \left(45-150 r_6\right)+15 r_6 \left(3 r_6-10\right)\right)\\
&\quad -r_2 r_4 \left(122 r_5^2+r_3 \left(15 r_6-4\right) r_5-60 r_6^2-750 r_6+r_1 \left(30 r_3 r_6+r_5 \left(15 r_6-4\right)\right)-60\right) \\
&\quad -r_2^2 \left(r_4 \left(45-150 r_6\right)+45 r_6\right)-30 \sqrt{r_1^2-4 r_2} \sqrt{r_3^2-4 r_4} \left(-5 r_4+r_2 \left(3 r_4-5\right)+3\right) r_6,\\
B_r &=3 \left(r_6 r_4^2+\left(r_6^2+1\right) r_4+r_6\right) r_1^2+\left(r_5 r_4^2+2 r_3 r_6^2\right) r_1+3 r_4 r_5^2-8 r_4 r_6^2+3 r_3^2 r_6-2 r_4^2 r_6\\
&\quad -12 r_4 r_6+r_3 r_4 r_5 r_6+r_2^2 \left(3 r_6 r_3^2+r_5 r_3+r_4 \left(3 r_5^2-12 r_6-8\right)-2 r_6\right)\\
&\quad +6 \sqrt{r_1^2-4 r_2} \sqrt{r_3^2-4 r_4} \left(r_6^2+r_2 r_4\right)+r_2 \left(3 \left(r_6^2+1\right) r_3^2+2 r_1 r_4 r_3+3 r_5^2-8 r_6^2\right)\\
&\quad +r_2 \left(r_4^2 \left(3 r_5^2-12 r_6-8\right)+r_1 r_5 r_6-12 r_6-12 r_4 \left(r_6^2-r_6+1\right)\right),\\
C_r &= -\left(r_4^2+r_6^2\right) r_1^2-r_4^2 r_5^2-r_3^2 r_6^2+r_4 r_6^2+r_4^2 r_6+r_2^2 \left(-r_3^2-r_5^2+r_4+r_6\right)+r_2 \left(r_4^2-6 r_6 r_4+r_6^2\right),
\ea\ee
a variable with a $t$ subscript is the same as the corresponding one with a $r$ subscript but with the polynomials $r_i$ replaced with the polynomials $t_i$, and
\begin{alignat*}{6}
r_1 &= a_1+a_2, &\quad r_2 &= a_1 a_2, &\quad r_3 &= a_3+a_4, &\quad r_4 &= a_3 a_4, &\quad r_5 &= a_5 +a_6, &\quad r_6 &= a_5 a_6,\\
t_1 &= a_1+a_4, &\quad t_2 &= a_1 a_4, &\quad t_3 &= a_2+a_3, &\quad t_4 &= a_2 a_3, &\quad t_5 &= a_5+a_6, &\quad t_6 &= a_5 a_6.
\end{alignat*}

We have numerically checked at several random kinematic points that the sum of all the diagrams vanishes for this process.

\subsubsection*{$\bullet$ $X^6$}

For the $X^a\bar{X}_b X^c\bar{X}_d X^e \bar{X}_f$ case we parameterize the momentum of the particles with the order appearing with $a_1,a_2,a_3,a_4,a_5,a_6$.  The conservation of momentum gives the two constraints
\be
\sum_{i=1}^6 a_i = \sum_{i=1}^6 \frac{1}{a_i}=0.
\ee
Then the contact term for the special case $a=c=e,~b=d=f$ is
\be
(S_1)^{aaa}_{bbb}=i \left(5A+30 (1-2a)B+45 (1-2a)^2 C \right) \delta^a_b,
\ee
where
\be\ba
A&=s_2 \left(t_2 \left(9 \left(8 s^2+125\right) u^2+9 \left(s^2+8\right)-638 s u+144 t_2 u^2\right)+144 s u\right)+144 s_2^2 t_2 u^2+144 s t_2 u,\\
B&=s_2 \left(8 t_2 \left(s^2-14 s u+18 u^2\right)+t_2^2 \left(-3 s u+9 u^2+8\right)+s (s (8 s u-3)+9 u)\right)\\
&\quad +s_2^2 t_2 \left(-3 s u+9 u^2+8\right)+s t_2 (s (8 s u-3)+9 u),\\
C&=4 s^2 t_2 \left(2 s u-t_2\right)-4 s_2^2 \left(s^2-2 t_2\right)+s_2 \left(8 s^3 u+t_2 \left(s^2-38 s u+8 t_2+45 u^2\right)\right),
\ea\ee
with $s_1,s_2,s_3$ symmetric polynomials in the $a_1,a_3,a_5$ variables and $t_1,t_2,t_3$ symmetric polynomials in the $a_2,a_4,a_6$ variables under the constraints
\be
s_1=-t_1=s, \quad s_3/s_2 = -t_3/t_2 = u.
\ee

We have numerically checked at several random kinematic points that the sum of all the diagrams vanishes for all processes involving six fields.

\subsubsection*{$\bullet$ Remaining six-point amplitudes}

We have also numerically checked that the scattering processes $X^a \bar{X}_b X^c \bar{X}_d Y^i Y^j$, $X^a \bar{X}_b Y^i Y^j Z Z$, $X^a \bar{X}_b Y^i Y^j Y^k Y^l$, $Y^i Y^j Z Z Z Z$, $Y^i Y^j Y^k Y^l Z Z$ and $Y^i Y^j Y^k Y^l Y^m Y^n$ vanish at several kinematic points.

\subsection{Comments on factorization of six-point amplitudes}

When we computed the scattering amplitudes above, we considered generic kinematics and showed that the five- and six-point scattering amplitudes vanish.  However, for some special kinematics, the amplitudes turn out to be non-vanishing.  This happens when one of the internal lines goes on-shell and the naive amplitude becomes infinite.  In this case, we need to keep track of the $i \epsilon$ prescription for the propagators.  The propagators can then be written as
\begin{equation}
  \frac i {p^2 - m^2 + i \epsilon} = \text{p.v.} \frac i {p^2 - m^2} + \pi \delta (p^2 - m^2),
\end{equation} by using the Sokhotskyi-Plemelj formula.

Using this formula, we can check that the principal value part cancels when summing all the diagrams, as for the case of generic kinematics (to show this rigorously we should scatter wave-packets instead of states with sharply defined momenta).  However, while the principal value part cancels in the sum of all the diagrams, in some cases the delta function survives.

It is easy to show that the five-point amplitude will always vanish, since, at the kinematic point where one internal propagator goes on-shell, the amplitude factorizes into a four-point and a three-point amplitude.  If the states are stable at tree-level, the on-shell three-point amplitude must vanish.  Therefore, the term containing the delta function in the expansion of the propagator also vanishes.

At six points, one can find a non-vanishing amplitude but with an extra delta function.  In the case of a $3 \to 3$ scattering with all particles having equal masses, the set of outgoing momenta is the same as the set of incoming momenta.  This should also hold for a $n \to n$ scattering, in which case the scattering amplitude contains $n$ delta functions.

\acknowledgments

We would like to thank M.~Spradlin for discussions and K.~Zarembo for correspondence.
CV would like to thank R.~Roiban for collaboration on related topics.  This research is supported in part by the US Department of Energy under contract DE-FG02-91ER40688 and by the US National Science Foundation under grant PHY-0643150 (CAREER/PECASE).

\appendix
\section{A toy model}
\label{sec:toy_model}

Before studying scattering in the $AdS_4 \times \mathbb{CP}^3$ sigma model which is the main focus of this paper, it worthwhile to look at a simpler model which has similar features.

The Tzitz\'eica model\footnote{This model is sometimes called Bullough-Dodd or Zhiber-Shabat model after the names of the authors who studied it in~\cite{Dodd:1977bi, Zhiber:1979}.} is an integrable model with one real bosonic field, which has three-point interactions, just like the $AdS_4 \times \mathbb{CP}^3$ sigma model.  This model, just like its cousin the sinh-Gordon model, could have been discovered by studying their tree-level scattering.  As described in~\cite{Goebel:1986na, Dorey:1996gd} one can show that the $n$-point tree-level scattering amplitudes can be made to vanish for $n>4$ by adding higher-point contact interactions to the Lagrangian.  Moreover, these interactions are tightly constrained by the requirement that the scattering amplitudes vanish.  By proceeding in this way one finds two models, the sinh-Gordon model, with Lagrangian
\begin{equation}
  \label{eq:sinh-Gordon_lagrangian}
  \mathcal{L} = \frac 1 2 (\partial \phi)^2 - \frac {m^2}{\beta^2} (\cosh (\beta \phi) - 1),
\end{equation} and the less-known Tzitz\'eica model with Lagrangian
\begin{align}
  \label{eq:Tzitzeica_model}
  \mathcal{L} = & \frac 1 2 (\partial \phi)^2 - \frac {m^2}{6 \beta^2} \left(e^{2 \beta \phi} + 2 e^{-\beta \phi} - 3\right) =\\
 = & \frac 1 2 (\partial \phi)^2 - \frac 1 2 m^2 \phi^2 - \frac 1 {3!} \lambda \phi^3 - \frac 1 {4!} \frac {3 \lambda^2}{m^2} \phi^4 + \dotso,  \label{eq:Tzitzeica_model2}
\end{align} where $\lambda = \beta m^2$.

Let us show the vanishing of the five-point scattering amplitude in the Tzitz\'eica model.  Instead of starting with the action in eq.~\eqref{eq:Tzitzeica_model}, we will start with some arbitrary couplings $\lambda_k$ in the action
\begin{equation}
  \mathcal{L} = \frac 1 2 (\partial \phi)^2 - \frac 1 2 m^2 \phi^2 - \sum_{k=3}^\infty \frac {\lambda_k} {k!} \phi^k.
\end{equation}
and show that $\lambda_3$, $\lambda_4$ and $\lambda_5$ should be related as in eq.~\eqref{eq:Tzitzeica_model2}.

For the five-point scattering process there are $26$ Feynman diagrams one needs to sum over, but they can all be obtained from the three five-point master topologies in Fig.~\ref{fig:5point} by permuting the external labels.  The diagrams are naturally organized into classes according to their parent topologies and the sum of all the diagrams belonging to a given class has the full permutation symmetry of the amplitude.

In the following we will use the light-cone momenta, which for an on-shell particle, $\varepsilon_i^2 =p_i^2 + m^2$, are defined by
\begin{equation}
  \label{eq:light-cone_momenta}
  a_i = \frac 1 m (\varepsilon_i + p_i), \quad \frac 1 {a_i} = \frac 1 m (\varepsilon_i - p_i).
\end{equation}  We will therefore express the amplitude in terms of the light-cone momenta $a_i$, instead of the usual momentum components $(\varepsilon_i, p_i)$.  If the theory is parity invariant, then $p_i \to - p_i$ or equivalently $a_i \to a_i^{-1}$ is a symmetry of the scattering amplitudes.  The Lorentz transformations act multiplicatively on the light-cone momenta $a_i \to t a_i$, where $t$ is a real number different from zero.

Now, if we use the fact that the sum of every class of Feynman diagrams which originate in the same master topology should be symmetric under the exchange of external momenta we conclude by using the fundamental theorem of symmetric polynomials that it can be represented in terms of elementary symmetric polynomials.  The elementary symmetric polynomials for $n$ variables $x_i$, with $i=1, \dotsc, n$ are defined by
\begin{equation}
  \label{eq:symmetric_polynomials}
  s_k = \sum_{1 \leq i_1 < \dotso < i_k \leq n} x_{i_1} \dotso x_{i_k},
\end{equation} for $k=1, \dotsc, n$.  So we will express the amplitude in terms of symmetric polynomials of light-cone momenta $a_i$.  The advantage of using the light-cone momenta $a_i$ is twofold: first, these variables solve the on-shell conditions and second, the momentum conservation imposes the constraints $s_1 = 0$ and $s_{n-1} = 0$ for $n$-point scattering.  In this language the parity symmetry acts by $s_k \to \tfrac {s_{n-k}}{s_n}$, with $1 \leq k \leq n$ and the convention $s_0 = 1$.

When reducing the sum of diagrams to symmetric polynomials and using the momentum conservation conditions $s_1 = 0$ and $s_{n-1} = 0$, the results simplify dramatically.  For example, for the Tzitz\'eica model, the sum of the $10$ diagrams corresponding to the second topology of Fig.~\ref{fig:5point} yields the result
\begin{equation}
  \label{eq:5point_second_top}
  \frac {i \lambda_3 \lambda_4}{m^2} \frac {2 s_2 s_3 - 5 s_5}{s_2 s_3 - s_5},
\end{equation} while the $15$ diagrams corresponding to the third topology of Fig.~\ref{fig:5point} yield
\begin{equation}
  \label{eq:5point_third_top}
  -\frac {i \lambda_3^3}{m^4} \frac {s_2 s_3 - 10 s_5}{s_2 s_3 - s_5}.
\end{equation}

Adding also the contribution of the five-point contact interaction, we get the following amplitude
\begin{equation}
  - i \lambda_5 + \frac {i \lambda_3 \lambda_4}{m^2} \frac {2 s_2 s_3 - 5 s_5}{s_2 s_3 - s_5} - \frac {i \lambda_3^3}{m^4} \frac {s_2 s_3 - 10 s_5}{s_2 s_3 - s_5},
\end{equation} which vanishes when
\begin{equation}
  \lambda_4 = 3 \frac {\lambda_3^2}{m^2}, \quad \lambda_5 = 5 \frac {\lambda_3^3}{m^4},
\end{equation} which are precisely the relations one obtains by expanding the Lagrangian in eq.~\eqref{eq:Tzitzeica_model}.

Let us perform a counting of the number of degrees of freedom for a $n$-point scattering amplitude.  The $n$ on-shell momenta have $2 n$ components from which we have to subtract $n$ on-shell constraints, two momentum conservation constraints and one constraint from Lorentz invariance (if the theory is Lorentz invariant).  Therefore, a $n$-point scattering process in two dimensions is characterized by $n-3$ parameters if the theory is Lorentz invariant and by $n-2$ parameters if the theory is not Lorentz invariant (in Sec.~\ref{sec:scattering_amplitudes} we deal with a theory which is not Lorentz invariant).

Let us first consider the case of theories which are not Lorentz invariant.  In this case, the $n-2$ parameters characterizing the kinematics can be taken to be some complex numbers $s_2, \dotsc, s_{n-2}, s_n$ (note that in the sequence above $s_1$ and $s_{n-1}$ are missing).  Starting with these $n-2$ numbers we form the degree $n$ equation
\begin{equation}
  (-x)^n + 0 (-x)^{n-1} + s_2 (-x)^{n-2} + \dotso + s_{n-2} (-x)^2 + 0 (-x) + s_n = 0.
\end{equation}  The $n$ solutions of this equation are the light-cone momenta $a_i$, with $i=1, \dotsc, n$.  In other words, the quantities $s_i$ are the elementary symmetric polynomials in the light-cone momenta $a_i$, by Vi\`ete's formulas.

In the case of a Lorentz invariant theory, a Lorentz transformation acting as $a_i \to t a_i$ for all $i$, transforms the quantities $s_k$ as $s_k \to t^k s_k$.  One can then eliminate a further parameter by this rescaling.

In Sec.~\ref{sec:scattering_amplitudes} we use the considerations in this appendix to obtain analytic expressions for some tree-level topologies.  However, there we do not have a full symmetry for some of the amplitudes, because the masses of the fields are not all equal.  Even in these cases, it will prove useful to use the partial symmetry in some of the particles to express their kinematics in terms of symmetric polynomials.  In the case of a theory with several particles of different masses, it might be helpful to study affine Toda theories as toy models, but we will not attempt this here.

\section{Commutation relations of the $so(2,3) \times su(4)$ algebra}
\label{sec:commutation_relations}

Here we present the non-vanishing commutation relations of the bosonic generators of the superalgebra $osp(6|4)$.  These commutation relations are taken from~\cite{Zarembo:2009au}, but we include them here to make the paper self-contained.

The commutation relations of the $u(3)$ subalgebra are:
\begin{alignat*}{3}
  [R_b^a, R_d^c] &= \delta_d^a R_b^c - \delta_b^c R_d^a, & \quad [R_b^a, R^c] &= \delta_b^a R^c - \delta_b^c R^a, & \quad [R_b^a, R_c] &= -\delta_b^a R_c + \delta_c^a R_b,\\
  [R^a, R_b] &= \delta_b^a (R_c^c - R) - R_b^a, & \quad [R, R^a] &= - R^a, & \quad [R, R_a] &= R_a.
\end{alignat*}

The commutation relations of the $so(1,3)$ subalgebra are:
\begin{equation*}
  [T_i, T_j] = \epsilon_{i j k} T_k, \quad [T_i, K_j] = \epsilon_{i j k} K_k, \quad [K_i, K_j] = - \epsilon_{i j k} T_k.
\end{equation*}

The remaining commutation relations for the $su(4)$ algebra are
\begin{alignat*}{4}
  [R_b^a, B^c] &= - \delta_b^c B^a, & \quad [R_a, B^b] &= \tfrac 1 2 \delta_a^b (J - M), & \quad [R, B^a] &= - B^a,\\
  [R_b^a, B_c] &= \delta_c^a B_b, & \quad [R^a, B_b] &= \tfrac 1 2 \delta_b^a(J + M) , & \quad [R, B_a] &= B_a,\\
  [R_b^a, J] &= \delta_b^a M, & \quad [R^a, J] &= B^a, & \quad [R_a, J] &= B_a,\\
  [R_b^a, M] &= \delta_b^a J, & \quad [R^a, M] &= -B^a, & \quad [R_a, M] &= B_a,\\
  [B^a, B_b] &= \delta_b^a R + R_b^a, & \quad [B^a, J] &= R^a, & \quad [B^a, M] &= R^a, &\\
  [B_a, J] &= R_a, & \quad [B_a, M] &= -R_a, & \quad [J, M] &= -2 R_a^a.
\end{alignat*}

The remaining commutation relations for the $so(2,3)$ algebra are
\begin{alignat*}{3}
  [K_i, L_j] &= - \delta_{i j} D, & \quad [T_i, L_j] &= \epsilon_{i j k} L_k, & \quad [K_i, D] &= -L_i,\\
  [L_i, L_j] &= -\epsilon_{ijk} T_k, & \quad [L_i, D] &= K_i.
\end{alignat*}

The conjugation properties of the generators are
\begin{alignat*}{4}
  (R_a^b)^\dagger &= R_b^a, & \quad (R^a)^\dagger &= R_a, & \quad R^\dagger &= R, & \quad (B^a)^\dagger &= - B_a,\\
  J^\dagger &= J, & \quad M^\dagger &= - M, & \quad D^\dagger &= -D,\\
  (L_i)^\dagger &= - L_i, & \quad (K_i)^\dagger &= - K_i, & \quad (T_i)^\dagger &= - T_i.
\end{alignat*}

Finally, the invariant bilinear form needed in eq.~\eqref{eq:sigma_model_action} is defined by (we only write the generators which we need in the expansion)
\begin{equation*}
  \Str B^a B_b = \delta_b^a, \quad \Str J^2 = -2, \quad \Str M^2 = 2, \quad \Str L_i L_j = \frac 1 2 \delta_{i j}, \quad \Str D^2 = -\frac 1 2.
\end{equation*}

\end{document}